\documentclass[a4paper,10pt]{article}

\usepackage{cmap}
\usepackage[T2A]{fontenc}
\usepackage[utf8]{inputenc}
\usepackage[english]{babel}
\usepackage{amsmath}
\usepackage{amssymb}
\usepackage{xcolor}
\usepackage{graphicx}
\usepackage{bm}
\usepackage{authblk}
\usepackage{nicefrac}
\usepackage{tabularray}
\usepackage{sidecap}
\usepackage[left=2cm,right=2cm,top=3cm,bottom=3cm]{geometry}

\sidecaptionvpos{figure}{c}

\author[1,2]{K.V.~Nikolaev}
\author[1,2]{L.R.~Muftakhova}
\author[3]{G.M.~Kuz’micheva}
\author[2,3]{Yu.N.~Malakhova}
\author[2]{A.V.~Rogachev}
\author[2]{N.N.~Novikova}
\author[2]{S.N.~Yakunin} 

\affil[1]{Moscow Institute of Physics and Technology, Dolgoprudny, Russia}
\affil[2]{National Research Center Kurchatov Institute, Moscow, Russia}
\affil[3]{MIREA -- Russian Technological University, Moscow, Russia}

\title{\bf Probing Langmuir monolayer self-assembly in condensed and collapsed phases: grazing incidence X-ray diffraction and X-ray standing waves studies}

\date{\today}

\begin{document}

\maketitle
\begin{abstract}

    Ce-induced effects on the self-assembly of arachidic acid monolayer was studied.
    The monolayers were formed on the cerium nitrate solution and studied in condensed and
    collapsed phases at two different temperatures $T = 21^\circ \rm{C}$ and $T = 23^\circ \rm{C}$.
    Grazing incidence X-ray diffraction and X-ray standing waves were applied to monitor in real
    time the changes in molecular packing.
    A new type of molecular organization was found,
    in which the monolayer maintains its lattice structure despite being overcompressed.
    Instead of forming crystalline aggregates or anisotropic cracking,
    the collapsed monolayer at $T=21^\circ \rm{C}$ appeared to be corrugated.
    The diffraction pattern for the monolayer in a new collapse mode is represented by the unclosed diffraction rings with maxima near the sample horizon.
    Theoretical formalism based on the distorted wave Born approximation was developed for quantitative analysis of the full 2D maps of diffraction scattering.
    The collapsed monolayer at $T = 21^\circ \rm{C}$ was shown to consist of crystalline flat domains, which are inclined from the horizontal position.
    The crystalline structure of a domain was identified as the pseudoherringbone packing mode.
\end{abstract}

\section{Introduction}

    Scientific interest in the self-assembly of amphiphilic trivalent lanthanide complexes is motivated by their photophysical properties,
    which can be exploited for molecular sensing~\cite{dossantos2008Recent}.
    The practical realization of such systems requires the assembly of the lanthanide complexes into nanostructures~\cite{wales2016surface}.
    One of the possible ways to create functional nanomaterials with predefined molecular architecture is the Langmuir-Blodgett technique~\cite{ariga2013lb}, 
    which involves the self-assembly of amphiphiles at the air-liquid interface into a ordered monolayer and subsequent transfer to a solid substrate.
    Charged headgroups of amphiphilic molecules are known to interact with metal ions when present in the liquid subphase~\cite{bosio1987reflectivity}.
    In this way, structural organization of the monolayer defines the performance of a nanodevice.
    Self-assembly is driven not only by the sort of amphiphiles and the thermodynamic conditions under which the monolayer is formed~\cite{kaganer1999structure},
    but can also be directed by interactions with ions in the subphase~\cite{barry2019chiral}.
    Therefore, further development of functional devices requires extensive and detailed studies of the structural organization of amphiphilic trivalent lanthanide complexes at liquid interfaces.
    
    There are two known scenarios of how the interaction with metal ions affects the structure of the Langmuir monolayer, 
    as reported in~\cite{kmetko2001effects} using divalent metal ions as an example. 
    The first, more common scenario is that the interaction between headgroups and metal ions leads to the formation of a disordered thin monolayer of metal ions underneath the organic monolayer, 
    which affects the compressibility of the monolayer, 
    resulting in structures similar to high-pressure solid-phase monolayers~\cite{kmetko2001effects}.
    The second scenario is the formation of the ionic monolayer ordered into a superstructure commensurate with the organic monolayer, 
    which was first reported in~\cite{leveiller1991crystallinity}.
    The study reports the formation of a Cd$^{2+}$ superstructure underneath an arachidic acid monolayer.
    The structure has a $2 \times 3$ cell commensurate with the organic monolayer structure. 
    The transition between these two scenarios of interaction of the organic monolayer with metal ions is known to be dependent on the pH of the subphase and the ion concentration. 
    In~\cite{dupres2003superlattice}, the interaction of behenic acid with a variety of divalent ions was studied, 
    and an ion concentration threshold was observed. Above this threshold, the ions formed a superlattice, and below it the ion layer was disordered. 
    Later studies~\cite{miller2016observation, miller2017atomic} report on forming ordered monolayers of lanthanide ions. 
    They tested a wide range of trivalent lanthanides. They found that the ordering depends not only on the molecules forming the Langmuir monolayer, 
    but also on the atomic number of a lanthanide. In addition, the crystalline structure of the ionic monolayer can be commensurate or incommensurate with the lattice of the organic molecules.
    
    Another interesting aspect is the collapse of Langmuir monolayers in the presence of metal ions. 
    The study in~\cite{zhang2020spontaneous} reports the spontaneous collapse of fatty acid monolayers in the presence of Ca$^{2+}$ ions.
    It has been shown both experimentally~\cite{vaknin2007ordering} and theoretically~\cite{lorenz2006atomistic} that the presence of Ca$^{2+}$ ions changes the structure of the collapsed fatty acid monolayer. 
    Even more interesting is that the collapsed structures show a high degree of order,
    namely the formation of inverted bi-layers. 
    The formation of such inverted bi-layers is also reported for lanthanides~\cite{nayak2022spontaneous}.
    Yet again, the structure of the collapsed layers is ion-specific.
    
    Thus, the interaction of fatty acids with lanthanide ions has shown complex and widely varying results. However, understanding the mechanisms of ion monolayer formation and its effect on collapse is relevant to the problems of functionalizing lanthanides into self-assembled nanostructures. 
    This article focuses on the self-assembly of arachidic acid (AA) monolayers in the presence of Ce$^{3+}$ ions and their structural reorganization while compressed beyond the collapse point. 
    We have studied the structure of fatty acid monolayers interacting with metal ions at the air-liquid interface at different surface pressures and temperatures.
    As the primary tool for monitoring the monolayer structure, 
    we used grazing incidence X-ray diffraction (GID) at the bending magnet synchrotron radiation source. 
    This choice is due to the sensitivity of GID to both translational and orientational lateral ordering of the monolayers. 
    In addition, simultaneously with GID experiments, 
    we performed the X-ray standing wave (XSW) measurements to monitor the vertical localization of Ce$^{3+}$ ions during the structural reorganization at the collapse.

\section{Results}

\subsection{Surface pressure-area isotherms}

    \begin{SCfigure}[][b!]
        \centering
        \includegraphics{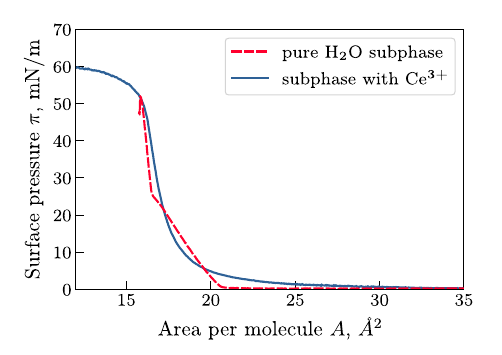}
        \caption{Surface pressure $\pi$ - area $A$ isotherms measured for the arachidic acid monolayer deposited on pure water (red line) and on cerium nitrate solution (blue line).
        The isotherm measured on pure water corresponds to a compression with two phase transitions: condensation and solidification. 
        The isotherm measured for a monolayer on a subphase containing Ce$^{3+}$ ions is characteristic of a phase transition from gas to solid with a compressibility of 2.2~m/mN, 
        bypassing the liquid phase.
        }
        \label{fig:iso}
    \end{SCfigure}

    The effect of Ce$^{3+}$ ions on the thermodynamic properties of the AA monolayer was investigated in our additional series of experiments using compression isotherm measurements.
    We have studied the behavior of the AA monolayer,
    formed on aqeous solution of 
    Ce(NO$_3$)$_3 \, \cdot \,$6H$_2$O salt
    and on pure water as a control. 
    All measurements were performed at a temperature of
    $T = 21^\circ$C.
    The typical $\pi$-$A$ isotherms are shown in Fig.~\ref{fig:iso}.
    Based on these experimental results,
    the key feature of the AA monolayer formed in the presence of Ce$^{3+}$ is the absence of a liquid phase,
    as evidenced by a high monolayer compressibility -- 2.2~m/mN.

    Our observations are consistent with the results reported in~\cite{sthoer2022la},
    where detailed studies of the binding interaction of lanthanide ions (La$^{3+}$ and Y$^{3+}$)
    with AA monolayers at submicromolar salt concentrations were reported.
    Note that similar changes in the isotherms were observed in~\cite{capistran2019effects},
    where the effect of Cu$^{2+}$ on the organization and morphology of the AA monolayer was examined.

    Generally, metal binding with carboxylate acid groups is recognized to promote the formation of a rigid monolayer with higher crystallinity and reduced molecular area~\cite{johann2001effect}.
    The interaction of divalent and trivalent metal ions with carboxylic groups of fatty acids is strongly dependent on such parameters as solution pH, 
    temperature, salt concentration. 
    The type of metal-carboxylate complex (ionic, unidentate, chelating bridging bidentate, polymeric tridentate)
    is determined to a large extent by metal ion specific nature,
    i.e. ion size, coordination structure, oxidation state in solution, etc.
    Large Ln$^{3+}$ ions with high coordination numbers exhibit rather intricate binding behavior~\cite{sthoer2022la}, 
    resulting in a wide variety of the lanthanide-carboxylate binding interactions. 
    According to the data reported in~\cite{sthoer2022la},
    at least three different coordination modes for carboxylate headgroups with lanthanide cations
    (La$^{3+}$ and Y$^{3+}$) have been identified at the air/water interface.
    Furthermore, the relative ratio between the different complexes was found to depend on the salt concentration in the subphase.
    Thus, the intricacies of the lanthanide complexation properties impose structural complications of the self-assembly processes in the fatty acid monolayers.
    The underlying mechanisms regulating these processes are the main focus of our studies.
    
\subsection{Brewster angle microscopy}

    \begin{figure}[b!]
        \centering
        \includegraphics{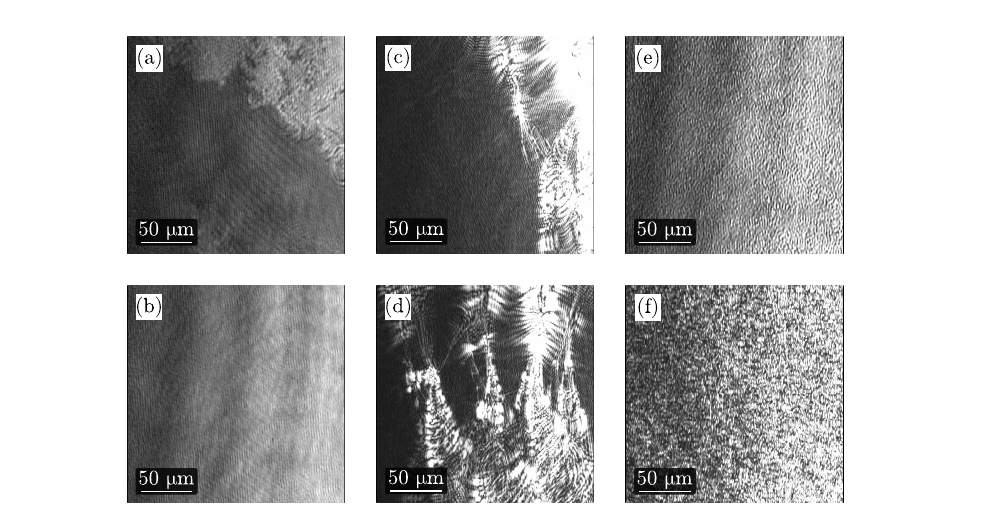}
        \caption{Brewster angle microscopy images of arachidic acid monolayers formed on the aqueous solution of the
                Ce(NO$_3$)$_3 \, \cdot \,$6H$_2$O salt.
                The images were taken at different stages of compression:
                (a) $\pi = 0.4$~mN/m;
                (b) $\pi = 35$~mN/m;
                (c) $\pi= 57$~mN/m;
                (d) $\pi = 58$~mN/m.
                The growth of the large crystalline-like aggregates can be seen in images (c) and (d) taken beyond the collapse point.
                The images (a)--(d) were taken at the temperature
                $T=23^\circ$C.
                The appearance of the mosaic-like collapse structure at the lower temperature
                $T=21^\circ$C
                is shown in (e) and (f).
                The image (e) was taken just beyond the collapse point, the image (f) was taken when the monolayer was further compressed.
        }
        \label{fig:bam}
    \end{figure}

    We performed BAM measurements of the AA monolayer formed on Ce$^{3+}$ solution
    to visualize the changes in morphological properties at different monolayer phases.
    The representative BAM images are shown in Fig.~\ref{fig:bam}.
    As can be seen, the BAM data obtained for the gas and solid phases indicate the formation of a uniform sheet-like layer.
    This kind of BAM images are usually recorded for fatty acid formed on a subphase with metal ions~\cite{capistran2019effects}.

    Ce-induced effects in the phase behavior of the AA monolayer were observed in BAM measurements in the collapse region (see Fig.~\ref{fig:bam}).
    We performed BAM studies of the AA monolayer at two different temperatures,
    $T = 21^\circ$C and $T = 23^\circ$C.
    BAM images taken at
    $T = 23^\circ$C
    showed the formation of large 3D crystalline-like aggregates.
    Such disordered structures are often observed in BAM studies of the collapse behavior of fatty acid monolayers.
    In contrast, the collapse BAM data obtained for the low temperature monolayer
    ($T=21^\circ$C)
    revealed the mosaic-like texture of the monolayer.
    As shown in Figs.~\ref{fig:bam}(e),~\ref{fig:bam}(f),
    the monolayer surface is divided into many patches,
    which can be identified as anisotropic domains that appear homogeneously across the monolayer upon compression.
    It should be emphasized that no growth of these domains into crystalline-like aggregates was observed upon further compression of the monolayer.

    Taken together,
    the presented BAM studies of AA monolayers formed on a subphase containing
    Ce$^{3+}$,
    revealed differences in the morphological properties of two observed collapsed modes.
    Thus, in our further discussion,
    the collapse with disordered morphology will be referred to as {\it collapse mode~1};
    whereas the collapse with specific mosaic-like morphology will be referred to as
    {\it collapse mode~2}.

    Note that similar results are reported in~\cite{hatta2003topological},
    where two qualitatively different collapse morphologies are also observed in fatty acid monolayers.
    They used phase contrast microscopy to study the collapse processes in the stearic acid monolayer as a function of the pH of the water subphase.
    According to the experimental data obtained in~\cite{hatta2003topological},
    the occurrence of collapse structures seen as bright random cracks was detected at $\text{pH}=7.8$,
    while at $\text{pH}=7.5$ a surface roughening collapse pattern was observed.

\subsection{Grazing incidence diffraction}

    \begin{table}[t!]
        \makebox[\linewidth][c]{
\begin{tblr}{ |l|| *{4}{c|}| *{2}{c|}| *{4}{c|}   }
    % Hat 1, ranges
    \hline
    \SetCell[r = 3]{c}{} & 
        \SetCell[c = 4]{c}{\bf range A} &&&& 
             \SetCell[c = 2]{c}{\bf range B} &&
                \SetCell[c = 4]{c}{\bf range C} 
                \\
    
    % Hat 2, peaks
    \hline
    &
        \SetCell[c = 2]{c}{ $(0 \, \nicefrac{1}{2})$ } &&
            \SetCell[c = 2]{c}{ $(\nicefrac{1}{3}\, \nicefrac{1}{2})$ } &&
                \SetCell[c = 2]{c}{ $(\nicefrac{2}{3}\, 0)$ } &&
                    \SetCell[c = 2]{c}{ $(1\,1)$, $(1\,\overline{1})$} &&
                        \SetCell[c = 2]{c}{ $(0\,2)$}
                        \\
    
    % Hat 3, controlled variables
    \hline
        & $Q_\parallel$,~\AA$^{-1}$ & $\delta$,~\AA$^{-1}$
        & $Q_\parallel$,~\AA$^{-1}$ & $\delta$,~\AA$^{-1}$
        & $Q_\parallel$,~\AA$^{-1}$ & $\delta$,~\AA$^{-1}$
        & $Q_\parallel$,~\AA$^{-1}$ & $\delta$,~\AA$^{-1}$
        & $Q_\parallel$,~\AA$^{-1}$ & $\delta$,~\AA$^{-1}$
        \\
        
    \hline \hline
    {\bf control}  
        & ---
            & ---
                & ---
                    & ---
                        & ---
                            & ---
                                & 1.496
                                    & 0.03
                                        & 1.496
                                            & 0.04  
                                                \\
    \hline
    {\bf series~1} 
        & 0.371
            & 0.01
                & 0.581
                    & 0.02
                        & ---
                            & ---
                                & 1.491
                                    & 0.03
                                        & 1.548
                                            & 0.02   
                                                \\
                                                
    \hline
    \SetCell[r = 3]{l}{{\bf series~2}$^*$}
        & 0.379
            & 0.01
                & 0.583
                    & 0.02
                        & ---
                            & ---
                                & 1.499 
                                    & 0.02 
                                        & 1.555
                                            & 0.03
                                                    \\
    
    \hline
        & \SetCell[r = 2]{c}{---} 
            & \SetCell[r = 2]{c}{---} 
                & \SetCell[r = 2]{c}{---} 
                    & \SetCell[r = 2]{c}{---} 
                        & \SetCell[r = 2]{c}{0.961} 
                            & \SetCell[r = 2]{c}{0.02}
                                & {\small 1.6$^{(11)}$}
                                    & {\small 0.07$^{(11)}$}
                                        & \SetCell[r = 2]{c}{1.479}
                                            & \SetCell[r = 2]{c}{0.01}
    \\
        & & & & & & 0.02
            & {\small  1.637$^{(1\overline1)}$}
                & { \small  0.01$^{(1\overline1)}$}
                    & &  \\
                    
    \hline
    {\bf series~3} 
        & 0.376 
            & 0.02 
                & \textasteriskcentered \textasteriskcentered 
                    & \textasteriskcentered \textasteriskcentered 
                        & ---
                            & ---
                                & 1.491 
                                    & 0.01 
                                        & 1.547 
                                            & 0.02   
                                                \\
    \hline
    \SetCell[r = 2]{l}{\bf series~4} 
        & \SetCell[r = 2]{c}{---}
            & \SetCell[r = 2]{c}{---}
                & \SetCell[r = 2]{c}{---}
                    & \SetCell[r = 2]{c}{---}
                        & \SetCell[r = 2]{c}{0.959}
                            & \SetCell[r=2]{c}{ 0.02}
                                &  {\small 1.595$^{(11)}$ }
                                    & {\small 0.02$^{(11)}$}
                                        &  \SetCell[r=2]{c}{1.476}
                                            & \SetCell[r=2]{c}{0.02} 
                                                \\
                                                
    & & & & & & 0.02
        & {\small 1.633$^{(1\overline1)}$ }
            & {\small 0.03$^{(1\overline1)}$}
                & & \\

    \hline
\end{tblr}
}
        \caption{Table of peak coordinates $Q_\parallel$ and corresponding full width at half maximum delta.
        The peaks are grouped by their $Q_\parallel$ in three ranges:
        range A: $Q_\parallel \sim [0.3,0.6]$~\AA$^{-1}$,
        range B: $Q_\parallel \sim 0.9$~\AA$^{-1}$ and range C: $Q_\parallel \sim [1.4,1.7]$~\AA$^{-1}$.
        *) The peaks of series 2 are assigned to the two coexisting phase modes,
        so this series is split into two rows in the table.
        **) Peak 
        $(\nicefrac{1}{3} \, \nicefrac{1}{2})$
        was not measured, but can be assumed to exist because
        $(0 \, \nicefrac{1}{2})$
        was measured and by analogy with series 1 and 2.
        These estimates are made by fitting the Gaussian profile to the data.
        For brevity, the uncertainties of these estimates are omitted.
        The values are given in the order of magnitude of the uncertainties.
        These uncertainties are further used to calculate the crystal structure errors in Table~\ref{tab:structure}.
        }
        \label{tab:peaks}
    \end{table}
    \begin{table}[t!]
        \center
        \begin{tblr}{|l| *{4}{|c} | *{2}{|c} | *{2}{|c}|}
    % Hat 2
    \hline 
    \SetCell[r = 3]{l}{}
        & \SetCell[c = 4, r = 2]{c}{\bf Organic lattice}
            &&&& \SetCell[c = 2, r = 2]{c}{\bf HC chain \\ ordering} 
                && \SetCell[c = 2, r = 2]{c}{\bf Super \\ structure} 
                    \\
    \\                            
    \hline
        & $a$,~\AA 
            & $b$,~\AA
                & $\gamma$,~deg
                    & A, \AA$^2\,$p.m.
                        & {tilt}
                            & {packing}
                                & inorganic
                                    & organic
                                        \\
                
    \hline\hline
    {\bf control}  
        & $4.85\pm0.02$
            & $8.40\pm0.02$
                & 90
                    & $20.4\pm0.1$
                        & NN
                            & ---
                                & ---
                                    & ---
                                        \\
    \hline
    {\bf series~1}
        & $4.93\pm0.03$
            & $8.12\pm0.02$
                & 90
                    & $20.0\pm0.2$
                        & NN
                            & ---
                                & $3\times2$
                                    & ---
                                        \\
    \hline
    \SetCell[r = 2]{l}{\bf series~2}
        & $4.90\pm0.03$
            & $8.12\pm0.02$
                & 90
                    & $19.8\pm0.2$
                        & NN
                            & ---
                                & $3\times2$
                                        & ---
                                            \\

    \hline
    &$4.4\pm0.2$
        & $8.5\pm0.1$
            & $88\pm5$
                & $19\pm1$
                    & no tilt
                        & PHB
                             & ---
                                & $3\times1$
                                    \\
    \hline
    {\bf series~3}
        & $4.93\pm0.02$
            & $8.12\pm0.02$
                & 90
                    & $20.0\pm0.1$
                        & NN
                            & ---
                                & $3\times2$
                                    & ---
                                        \\
    \hline
    {\bf series~4}
        & $4.38\pm0.02$
            & $8.52\pm0.02$
                & $88.3\pm0.4$
                    & $18.6\pm0.1$
                        & no tilt
                            & PHB
                                 & ---
                                    & $3\times1$
                                        \\
    \hline
\end{tblr}
        \caption{
            List of lattice parameters
            (rectangular centered cell assumed),
            phase modes corresponding to the orientational ordering of the hydrocarbon chains and superstructures as deduced from the GID patterns in Fig.~\ref{fig:all}.
        }
        \label{tab:structure}
    \end{table}

\begin{SCfigure}[][t!]
        \centering
        \includegraphics{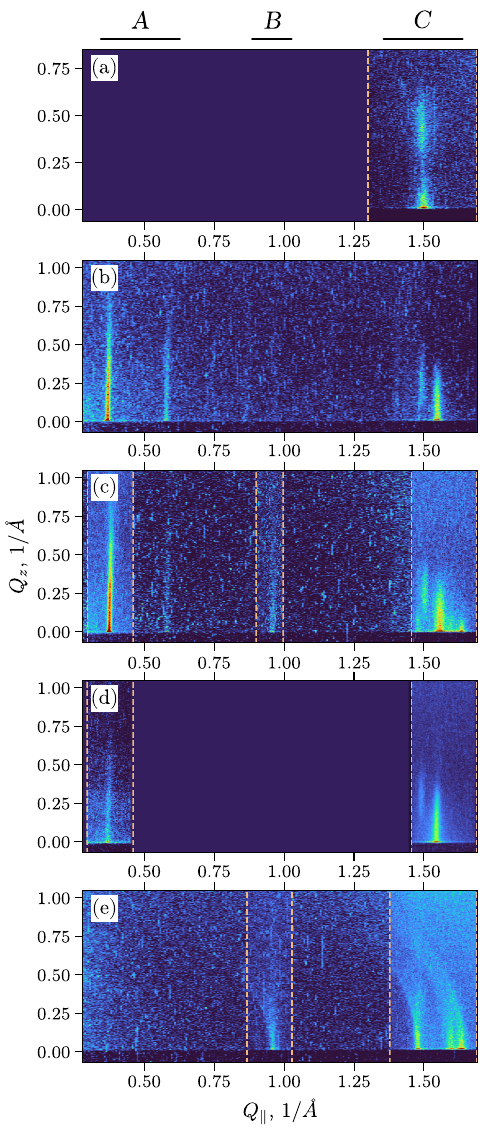}
        \caption{   Summary of all GID pattern measurements.
                    (a) control series
                    (b) -- (e) series 1 through 4, respectively.
                    Each pattern is collected from all measurements made on a selected,
                    most representative sample in the series.
                    Some measurements are taken in the different $Q_\parallel$ range.
                    This is indicated by the vertical dashed lines.
                    The diffraction peaks are grouped by their $Q_\parallel$ coordinate into three ranges:
                    A, B and C.
                }
        \label{fig:all}
    \end{SCfigure}

    Five series of GID measurements were performed in our studies.
    Each series was designed to study the structure of the AA monolayer at different surface pressure and temperature.
    There is one control series in which the AA monolayers were deposited on pure water subphase and four series in which AA monolayers were formed on a
    20~mM Ce(NO$_3$)$_3 \, \cdot \,$6H$_2$O salt solution. 
    Within each series, except the control, multiple samples of the monolayer were prepared and measured. This was done to ensure reproducibility. 
    The results are shown in Fig.~\ref{fig:all}. 
    The measurements from the most representative samples are selected to be shown.
    From these selected samples,
    all collected data were integrated into a single GID pattern to represent each series.
    Some measurements were taken with different resolution or different scan area.
    These measurements are superimposed on the GID pattern, 
    which is indicated by vertical dashed lines.
    It is important to note that for some measurements the GID pattern has evolved over time. 
    Since the data is integrated over all measurements, different stages of structural change are captured as "motion blur" in Fig.~\ref{fig:all}. For such cases, the time evolution is shown
    in Figs.~\ref{fig:series2}, \ref{fig:series3} and \ref{fig:series4}. 
    
    We group the observed diffraction peaks into three ranges according to their $Q_\parallel$ coordinate. 
    Range A: low spatial frequency  $Q_\parallel \sim [0.3,0.6]$~\AA$^{-1}$.
    Range B: medium spatial frequency  $Q_\parallel \sim 0.9$~\AA$^{-1}$.
    Range C: high spatial frequency $Q_\parallel \sim [1.4,1.7]$~\AA$^{-1}$;
    this is the range in which a typical diffraction pattern from a Langmuir monolayer is observed. 
    The peak coordinates and corresponding full width at half maximums are listed in Table~\ref{tab:peaks}.
    Finally, the crystal structure is inferred from the GID patterns for each series. The results are shown in Table~\ref{tab:structure}.
    Note that throughout this article we use the rectangular centered unit cell with the corresponding peak index notation.

\subsubsection*{\bf Control series: monolayer on pure water}
    An AA monolayer was formed on the surface of pure water and compressed to a surface pressure $\pi = 20$~mN/m at temperature $T = 21^\circ {\rm C}$.
    In this series of measurements [see Fig.~\ref{fig:all}(a)] a typical GID pattern for a Langmuir monolayer was observed: a triplet of peaks $(02)$, $(11)$, and $(1\overline1)$.
    The brighter peak at the horizon of the monolayer is the $(02)$ peak,
    the peak above the horizon is a degenerate peak,
    in which two peaks $(11)$ and $(1\overline 1)$ are superimposed.
    The arrangement of the peaks on top of each other indicates a hexagonal lattice:
    $b = \sqrt{3}a$, $\gamma = 90^\circ$.
    %TODO: add: we use rectangular centered parameters
    The tails of the fatty  molecules are inclined with respect to the surface and are tilted in the direction of the nearest neighbor which corresponds to the NN tilt phase mode.
    We infer this from the arrangement of peaks $(02)$ on the horizon and $(11)$ directly above, 
    which is characteristic of the NN tilt~\cite{kaganer1999structure}.

\subsubsection*{\bf Series~1: solid-phase monolayer at T~=~23$^\circ$C } 
    The monolayer has been compressed to a surface pressure $\pi = 20$~mN/m
    at $T = 23^\circ{\rm C}$.
    GID data were collected continuously over a 13~hour period in multiple exposures.
    The integrated GID pattern is shown in Fig.~\ref{fig:all}(b). 
    The surface pressure was kept constant at $\pi = 20$~mN/m throughout all exposures.
    The GID patterns in all exposures are equivalent down to the measurement noise.
    Thus, the structure is unchanged throughout the 13~hour period.
    This is quite remarkable because it indicates that the monolayer was not damaged by the radiation.
    Note that the radiation damage is a typical problem in the synchrotron studies of Langmuir monolayers, see e.g.~\cite{strzalka2004resonant}.
    Here, we would like to highlight the importance of measuring the GID at lower intensities and with longer exposures,
    as the self-assembly process may be slower than the degradation induced by radiation.
    From this point of view,
    measuring on a bending magnet beamline can be advantageous compared to brighter sources such as insertion devices.

    In range C we again see a $(02)$ peak at the horizon
    and a degenerated peak $(11)$, $(1\overline1)$ above the horizon.
    However, the GID pattern observed in the control series,
    the $(02)$ peak is shifted horizontally to higher frequencies.
    This implies a lattice with tighter packing,
    compared to that of the AA monolayer on pure water:
    the surface area occupied by a molecule is reduced;
    20~\AA$^2$ for series~1 versus 20.4~\AA$^2$ for control.
    The lattice is not hexagonal ($b \neq \sqrt{3}a$),
    however the lattice angle is the same: $\gamma = 90^\circ$.
    The structure has NN tilt.

    Of particular interest are the bright peaks in the range~A.
    We attribute these peaks to diffraction at a layer formed by Ce$^{3+}$ ions adsorbed at the bottom of the monolayer on the surface of the subphase.
    This assumption is confirmed by two observations.
    First, the low spatial frequencies $Q_\parallel$ of the peaks imply a larger spatial period compared to the organic lattice.
    The $Q_\parallel$ coordinates of these peaks correspond to the fractional indices 
    $(0 \, \nicefrac{1}{2})$ and 
    $(\nicefrac{1}{3} \, \nicefrac{1}{2})$
    of the reciprocal unit cell of the AA monolayer.
    This implies that Ce$^{3+}$ ions form a superstructure underneath the AA monolayer.
    The unit cell of such a superstructure has a size of
    $3 \times 2$
    unit cells of the AA molecular lattice.
    Second, the vertical size of the inorganic lattice peaks at
    ($Q_{z,{\rm max}} \sim 0.8$ \AA$^{-1}$)
    is significantly larger than that of the organic lattice peaks
    ($Q_{z,{\rm max}} \sim 0.4$ \AA$^{-1}$),
    indicating a greater vertical spatial localization of Ce$^{3+}$ ions compared to 
    that of the AA molecules.
    The formation of the Ce$^{3+}$ layer at the subphase interface below the monolayer is also supported by the XSW measurements, as will be discussed in the next section.
    We will refer to the superstructure of Ce$^{3+}$ ions as the {\it inorganic superstructure}. 

\subsubsection*{\bf Series~2: solid-phase monolayer at T = 21$^\circ$C}

\begin{figure}[t!]
    \centering
    \includegraphics{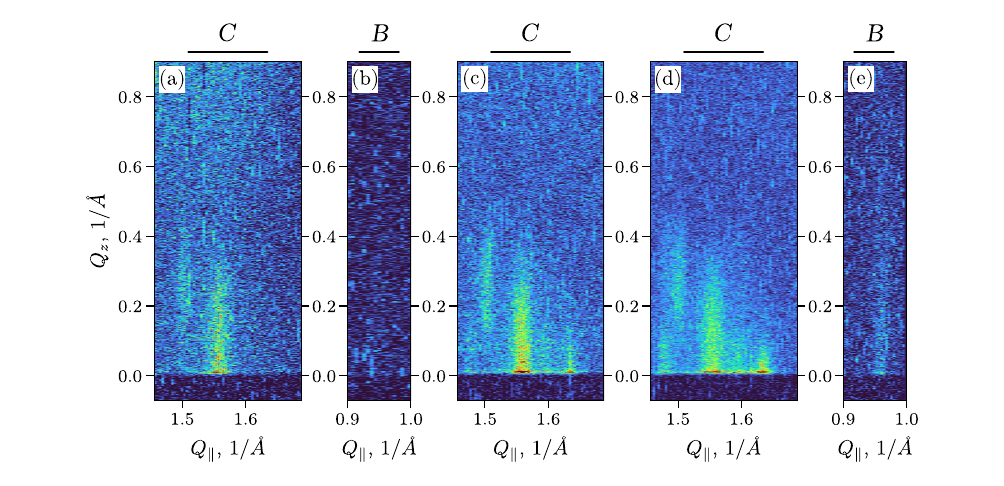}
    \caption{   Time evolution of the GID pattern in series 2.
                The start and end times of the measurement in terms of hours elapsed
                after monolayer compression:
                (a) 2 -- 3 hrs.
                (b) 3 -- 7 hrs.
                (c) 7 --  11 hrs.
                (d) 11 -- 18 hrs.
                (e) 21 -- 22 hrs.
                Patterns are measured in different $Q_\parallel$ ranges,
                which are indicated above the frames.
                Such a change in the GID pattern indicates that the monolayer was initially in a solid phase with an NN tilt and an inorganic superstructure,
                after which there was a coexistence of solid phase modes with an NN tilt and a PHB ordering.
            }
    \label{fig:series2}
\end{figure}

    The monolayer was compressed to a surface pressure of $\pi = 20$~mN/m as in series~1.
    However, this time the samples was prepared at a lower temperature of $T=21^\circ$C.
    All GID measurements from a selected sample are integrated and shown in Fig.~\ref{fig:all}(c).
    The total duration of the measurements on this sample was 22 hours.
    The surface pressure was kept constant throughout the measurements.

    The GID pattern changed during the measurements, its evolution is shown in Fig.~\ref{fig:series2}.
    Similar to series~1, it exhibits a typical GID pattern, characteristic of an NN tilt, in the range C [see Fig.~\ref{fig:series2}(a)].
    This pattern was first observed 3 hours after compression.
    The full range measurement shown in Fig.~\ref{fig:all}(c) was performed over the next 4 hours.
    A part of this measurement is also shown in Fig.~\ref{fig:series2}(b) to emphasize that no peaks were detected in the B region during this period. The GID pattern of the inorganic superstructure was also observed [see peaks in range A in Fig.~\ref{fig:all}(c)].
    The change in GID pattern was first observed 11 hours after compression [see~Fig.~\ref{fig:series2}(c)].
    It evolved to a final GID pattern [see~Fig.~\ref{fig:series2}(d)] during another 7 hours of measurements.
    In this pattern we observed five peaks in the range~C.
    Three more peaks appear in addition to the NN pattern.
    We interpret the pattern in~Fig.~\ref{fig:series2}(d) as diffraction from two coexisting packing modes.
    The additional peaks are another triplet of $(11)$, $(1\overline1)$ and $(02)$, corresponding to a new packing mode.
    Since we can distinguish all three peaks at different $Q_\parallel$,
    the lattice in this mode should have an oblique unit cell.
    We can also conclude that there is no tilt of the AA molecules, since all three peaks are on the horizon of the sample.

    Finally, 22 hours after the compression, we detected another peak at $Q_\parallel = 0.961$~\AA$^{-1}$ in range B.
    We ascribe this peak with diffraction on a superstructure corresponding to the lattice of a new packing mode.
    The peak is identified by the fractional indices
    $(\nicefrac{2}{3}\, 0)$.
    In contrast to the diffraction on the inorganic superstructure, this peak is considerably smaller in the $Q_z$ direction with
    $Q_{z,{\rm max}} \sim 0.4$~\AA$^{-1}$.
    On the other hand, it is comparable in $Q_{z,{\rm max}}$ to the peaks of the organic structure in range~C.
    Therefore, this is not a diffraction at the Ce$^{3+}$ lattice,
    but diffraction on a superstructure composed of the AA molecules -- {\it organic superstructure}.
    It can appear in the Langmuir monolayers due to the azimuthal ordering of the hydrocarbon chains of the molecules~\cite{dupres2002evidence,kuzmenko1998packing}.
    Molecules of the monolayer can assume a denser packing due to a specific mutual orientation of the hydrocarbon chains of neighboring molecules,
    that is called herringbone (HB) packing -- an ordering when the azimuth angle between the hydrocarbon chains is 90$^\circ$, and pseudoherringbone (PHB) packing when this angle is about 40$^\circ$~\cite{fradin1998microscopic}.
    Similar organic superstructure peaks in the GID have also been reported for AA films in~\cite{peng2000superstructures},
    but there the Langmuir-Blodgett multilayers on solid substrate were studied.
    They report a superstructure with $5 \times 1$ cell and an HB arrangement.
    The $(\nicefrac{2}{3} \, 0)$ peak in our study can be ascribed to a $3 \times 1$ superstructure and a PHB arrangement. It should be emphasized that in our experiments we were able to monitor in real time the gradual transition from the NN tilt mode to PHB packing. 
    
    To summarize the series 2,
    the monolayer was initially observed in a solid phase with rectangular lattice, NN tilt and inorganic superstructure, identical to series~1.
    However, in contrast to series~1, a new packing mode was observed 11 hours after compression.
    We attribute this to the PHB ordering.
    This packing mode coexists with an initial phase.
    
\subsubsection*{\bf Series~3: collapse mode~1}

    \begin{figure}[t!]
        \centering
        \includegraphics{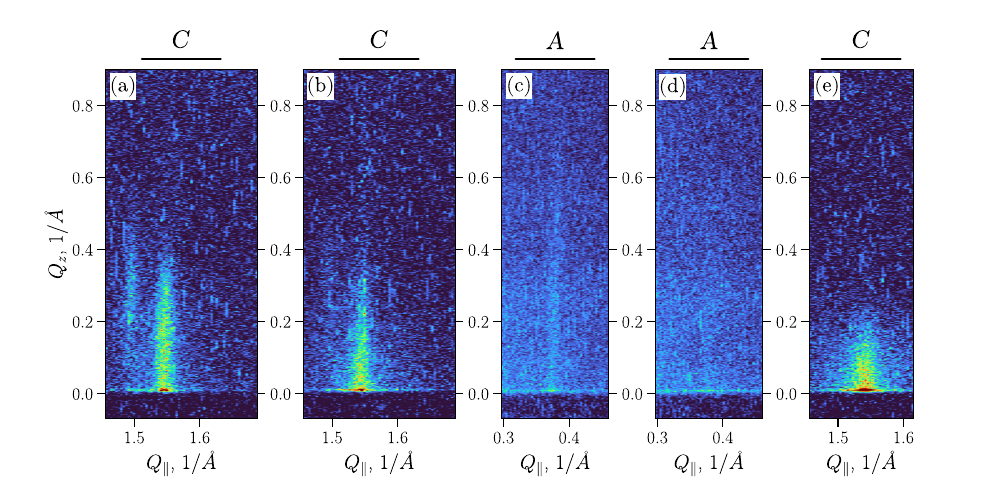}
        \caption{   Time evolution of the GID pattern in series 3.
                    The start and end times of the measurement in terms of hours elapsed
                    after monolayer compression:
                    (a)  1 -- 2 hrs.
                    (b)  18 -- 19 hrs.
                    (c)  21 -- 24 hrs.
                    (d)  26 -- 29~hrs.
                    (e)~40 -- 41 hrs.
                    Patterns are measured in different $Q_\parallel$ ranges,
                    which are indicated above the frames.
                    This evolution of the GID pattern corresponds to the gradual collapse of the monolayer starting from a structure with NN tilt phase and inorganic superlattice to a system with only short range order and no superstructure.
                    }
        \label{fig:series3}
    \end{figure}

    The monolayer samples in this series have been compressed beyond the collapse point.
    An integrated GID pattern is shown in Fig.~\ref{fig:all}(d).
    GID measurements for this sample were taken over a 41-hour period.
    The temperature is $T=23^\circ$C.
    The surface pressure $\pi = 52$~mN/m was reached during compression,
    after which the surface area was kept constant throughout the measurements.
    
    In the first GID pattern measured just after compression, we again see a GID pattern characteristic of the NN tilt mode
    [see Fig.~\ref{fig:series3}(a)].
    The GID pattern of the inorganic superstructure was also detected after compression, as seen in the integrated pattern in Fig.~\ref{fig:all}(d): note the peak in the range~A.
    However, the GID pattern has started to degrade,
    [Fig.~\ref{fig:series3}(b)] measured 19 hours after compression. 
    In addition, the peak from the inorganic superstructure is still visible 24 hours after compression.
    In the measurement 5 hours later, the peak from the inorganic superstructure has disappeared.
    Subsequently, the GID pattern of the monolayer in the range~C also degraded, as seen in Fig.~\ref{fig:series3}(e).

    For the inorganic superstructure, only the $(0 \, \nicefrac{1}{2})$ peak was monitored to save measurement time for resolving the structural change of the monolayer.
    Since it is attributed to an inorganic superstructure,
    the $(\nicefrac{1}{3} \, \nicefrac{1}{2})$ is also very likely to exist.

    Thus, in series~3, the monolayer was compressed beyond the collapse point.
    Initially, the structure of the monolayer was similar to that of a solid phase monolayer as in series~1 and~2,
    but the structure began to degrade,
    and 41 hours after compression, only the peak corresponding to the short-range order was present,
    indicating the lateral disordering of the monolayer.
    The inorganic superstructure has also degraded as a result of the collapse.

\subsubsection*{\bf Series~4: collapse mode~2}
    \begin{figure}[t!]
        \centering
        \includegraphics{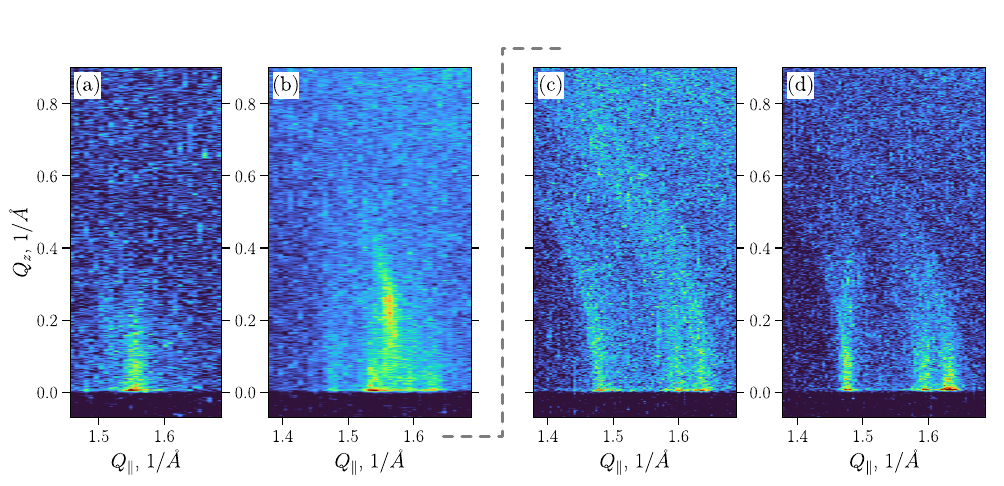}
        \caption{The time evolution of the GID pattern in series~4.
                 Note that the data were measured from two different monolayer samples deposited under the same conditions.
                 (a), (b) is for the first sample,
                 (c), (d) is for the second one.
                 This is indicated by the dashed line.
                 The start and end times of the measurement in terms of hours elapsed after monolayer compression:
                 (a) 0 -- 2 hrs.
                 (b) 3 -- 11 hrs.
                 (c) 3 -- 5 hrs.
                 (d) 22 -- 24 hrs.
                 }
        \label{fig:series4}
    \end{figure}

    In this series,
    a new collapse mode of the AA monolayer was studied.
    The integrated GID pattern is shown in Fig.~\ref{fig:all}(e).
    The monolayer has been compressed beyond the collapse point at $\pi = 52$~mN/m,
    after which the compression was stopped.
    The monolayer was studied at temperature $T = 21^\circ$C.

    The GID pattern in Fig.~\ref{fig:all}(e)
    is unusual for diffraction on a 2D~polycrystalline system:
    the Bragg rods are rounded.
    The rounding of the peaks is exactly on the circle in reciprocal space and around its origin.
    In this sense,
    these peaks are similar to the Scherrer rings characteristic of diffraction on a 3D polycrystalline material.
    In our case,
    these diffraction rings are not closed.
    One can see the rounded rods instead.
    This suggests that collapse mode~2 is an intermediate state of the monolayer between a 3D and a 2D powder.  
    In fact, the Langmuir film retains high degree of structural organization because we still observe the GID pattern characteristic of an ordered monolayer, namely a monolayer with PHB hydrocarbon chain ordering.
    Moreover, the positions of peaks $(11)$, $(1\overline1)$ and $(02)$ are the same as in series~2 
    [see Figs.~\ref{fig:all}(c),~\ref{fig:all}(e) and Table~\ref{tab:peaks}].
    Their $Q_\parallel$ coordinates are within 1\% deviation between series 2 and 4.
    There is also a peak in range~B [see Fig.~\ref{fig:all}(e)], again suggesting the PHB packing mode.
    However, these 2D domains are oriented in 3D with their normal deviating from the normal of the subphase.
    Presumably, in the collapse mode~2, the structure can be ascribed to a corrugated monolayer, as the domains are randomly oriented with a preferential orientation in the subphase plane.

    The time evolution of the monolayer in the collapse mode~2 is shown in Fig.~\ref{fig:series4}.
    In this series of measurements, the monolayers underwent a rapid structural change compared to the measurement timescale.
    We measured both XSW and GID.
    To do this, we prepared several monolayer samples in collapse mode~2.
    In Fig.~\ref{fig:series4} we show GID data from two different samples.
    For both samples all controllable conditions were the same.
    On the first sample, we performed the GID measurements just after compression.
    The GID pattern measured within the first 2~hours is shown in Fig.~\ref{fig:series4}(a),
    and the data collected between 3 and 11~hours after compression are shown in Fig.~\ref{fig:series4}(b).
    On the second sample,
    we first performed the XSW measurements,
    which will be discussed in the next section,
    starting at 3 hours after compression we performed several GID measurements in between XSW scans.
    This experimental strategy allowed us to analyze the dynamics of the Ce distribution during the structural change in collapse mode~2.
    The GID pattern collected between 3 and 5~hours after compression is shown in Fig.~\ref{fig:series4}(c)
    and between 22 and 24 hours is shown in Fig.~\ref{fig:series4}(d).

    As in all other measurements, the monolayer initially formed into a structure with the NN tilt packing mode [see Fig.~\ref{fig:series4}(a)].
    However, as the measurements continued, the GID revealed the structural change. 
    Fig.~\ref{fig:series4}(b) shows a rich diffraction pattern, which we assume to be a superposition of two patterns: 
    three lower intensity peaks of PHB located exactly at the same $Q_\parallel$ coordinates (within 1\% accuracy) as the PHB peaks in series~2; 
    and two brighter peaks similar to a typical picture of the NN tilt, but with $Q_\parallel$ coordinates different from those in all other series.
    Also, unlike all other measured NN tilt patterns, the peak above the horizon is to the right of the peak on the horizon.
    Furthermore, one can see that the bright peak above the horizon is slightly rounded.

    In the pattern measured on another sample [see Fig.~\ref{fig:series4}(c)], the roundness of the peaks is clearly visible. There are no other peaks associated with the NN tilt mode in this pattern. The peaks are on circles centered at the origin of the reciprocal space. The pattern remained unchanged while the surface pressure was kept constant by shrinking the surface area. This continued until the minimum position of the trough barrier was reached 5~hours after the initial compression. After that, the GID pattern changed again. The pattern in Fig.~\ref{fig:series4}(c), obtained 24 hours after compression, has a subtle difference. The peaks in this pattern appear to be partly rounded and partly straight. We assume that this GID pattern is the incoherent sum of two patterns: one with rounded peaks and another with straight peaks. However, to confirm this assumption, numerical simulations of evanescent X-ray diffraction are required. This will be considered further, but first we would like to point out another noteworthy observation of series~4. Namely, there are no inorganic superstructure peaks [see Fig.~\ref{fig:all}(e)]. This immediately raises the question of the localization of Ce in the collapse mode~2. This question can be answered with the help of X-ray standing waves.
    
\subsection{X-ray standing waves}

    \begin{figure}[t]
        \centering
        \includegraphics{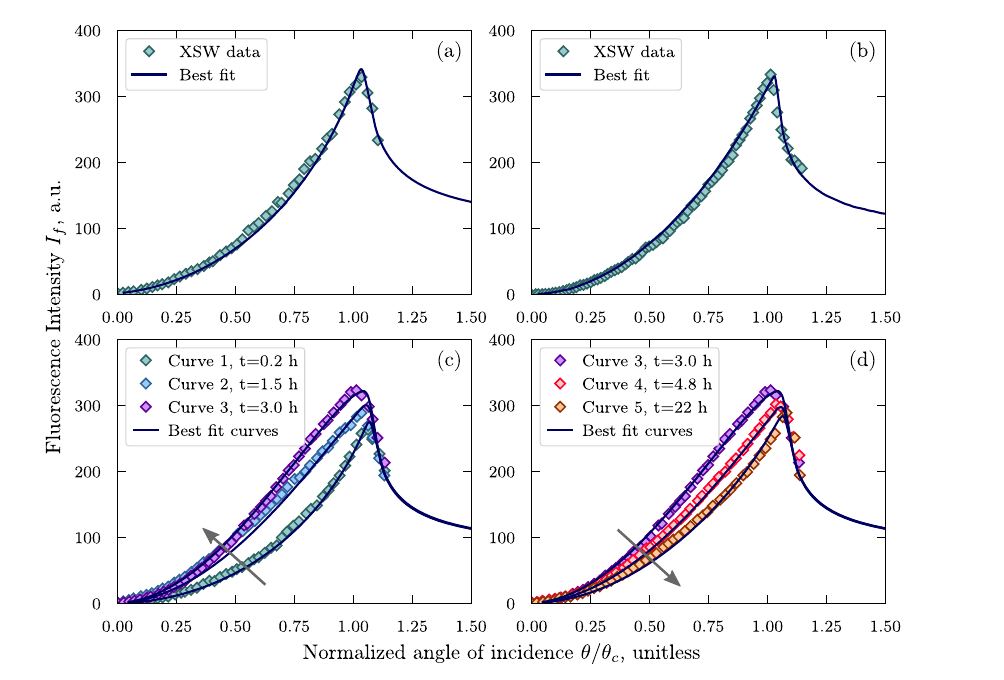}
        \caption{X-ray standing wave measurements: fluorescence intensity characteristic of Ce measured with respect to the angle of incidence $\theta$. 
        (a) Measured for the solid phase (one of many typical curves measured in series 1 and 2). 
        (b) Measured for the collapse mode~1 (series 3). (c), (d) The curves measured for the collapse mode~2 (series 4). 
        The curves are measured at different times on the single monolayer. 
        The time elapsed after the monolayer compression is indicated in the legend of the figure. 
        The arrows indicate the direction of change in the slope of the curve over time.}
        \label{fig:XSW}
    \end{figure}

    We applied X-ray standing wave (XSW) technique to monitor the changes in the
    location of cerium atoms during the self-assembling reorganization of the AA monolayer.
    In XSW studies at liquid interfaces, the intensity of secondary radiation
    (e.g., characteristic fluorescence),
    excited by the incident X-rays, is measured as a
    function of the incident angle in the subcritical region
    $0 < \theta <\theta_c$.
    The angular dependence
    of fluorescence intensity is known to be highly sensitive to the position of the atoms emitting
    secondary radiation signal,
    that allows to locate the atoms by the fitting the
    corresponding fluorescence curve.
    For quantitative analysis of XSW data
    the fluorescence intensity curves are usually simulated using the Parratt equations~\cite{parratt1954surface} or optical transfer matrices (see~\cite{gibaud2000reflectivity, nikolaev2023grazing} among others).

    In our experiments, the intensity of Ce-fluorescence was measured for the solid phase and the two collapse modes (Fig.~\ref{fig:XSW}). 
    The XSW measurements for collapse mode~1 were performed at an energy of $E = 5.8$~keV, 
    the other Ce-fluorescence curves were measured at $E = 13$~keV.

    The XSW measurements for the solid-phase monolayer were carried out 21 hours after the
    AA monolayer was compressed to $\pi = 20$~mN/m.
    Fig.~\ref{fig:XSW}(a) shows the angular dependence of integrated intensity
    for Ce~$L_\alpha$ peak obtained in these measurements.
    From this figure, the intensity of the Ce-fluorescence increases dramatically from zero
    at $\theta = 0^\circ$ and reaches the sharp maximum near the critical angle $\theta = \theta_c$.
    This is a typical behavior,
    which is observed for the angular dependence of the fluorescence intensity
    from atoms distributed in the near-surface region~\cite{zheludeva2013biomembrane}.
    Theoretical simulations have shown that there is a good
    agreement between the calculations and the experimental data,
    for the solid-phase monolayer,
    which can be obtained within the frame of a two-layer model:
    the Ce-free top layer (AA~monolayer)
    and the Ce-containing bottom layer. 
    Both layers were considered to be homogeneous.
    The theoretical Ce-fluorescence curve in Fig.~\ref{fig:XSW}(a) was calculated for the
    case when Ce atoms are present in a 5~\AA{} thick layer under the top layer with a
    thickness of 27~\AA{}.
    Based on these data we can suggest that Ce atoms formed a single
    layer underneath the AA monolayer,
    that is a quite reasonable arrangement of Ce atoms for a solid-phase monolayer.
    
    The XSW data for collapse mode~1 are presented in Fig.~\ref{fig:XSW}(b).
    The monolayer was compressed to a surface pressure of $\pi = 55$~mN/m,
    after that layer area was held constant.
    The angular dependence of the Ce-fluorescence intensity was recorded 1.5~hours
    after the AA monolayer was compressed.
    As can be seen, this experimental Ce-curve is sharply peaked at the critical angle $\theta_c$
    and is very similar to that recorded for the solid-phase monolayer.
    This observation is quite remarkable indicating that in the collapse
    mode~1, Ce atoms are located in the near-surface region.

    The arrangement of Ce atoms in collapse mode~2 appeared to be more complicated.
    XSW data, obtained for collapse mode~2,
    are shown in Figs.~\ref{fig:XSW}(c),~\ref{fig:XSW}(d).
    These measurements were taken during GID series~4.
    A total of six XSW curves were recorded at different times elapsed after compression of monolayer beyond the collapse point ($\pi > 52$~mN/m).
    The time of measurement is marked in the legend of Figs.~\ref{fig:XSW}(c),~\ref{fig:XSW}(d).
    As can be seen the shape of these curves markedly evolved over time.
    The first Ce-fluorescence curve exhibited the behavior,
    which is very similar to that observed for solid-phase monolayer; but the other Ce-fluorescence curves appeared to be significantly different.
    The most obvious difference is the pronounced convexity near the critical angle.
    To interpret the observed changes in the Ce-fluorescence curves we hypothesized that some of cerium atoms in collapsed film went upwards.
    In such a case the experimental Ce-fluorescence can be fitted as the weighted sum of the angular dependencies of the fluorescence intensity from cerium atoms located above  the liquid surface and from cerium atoms, 
    which are present underneath the AA monolayer.
    The shape of the resulting Ce-fluorescence intensity depends mainly on three parameters: 
    two weighting coefficients ($C_1$ and $C_2$) and the thickness of the Ce distribution above the liquid surface.
    The best-fit curves in Figs.~\ref{fig:XSW}(c),~\ref{fig:XSW}(d),
    were obtained by varying the weight coefficients while keeping the thickness of the Ce distribution above the liquid surface constant at 55~\AA{}.
    The parameters for the Ce layer located underneath AA film were fixed to those determined from the analysis of the first XSW curve, recorded in this series.
    Note, that the best-fit parameters for the first XSW curve coincide with the those,
    obtained for solid-phase AA monolayer.

    To understand the general trends in the changes in the location of cerium ions
    it is convenient to use the weight coefficient ratio $C_{12}=C_1/C_2$,
    that allows to estimate the fractional amount of the cerium ions,
    located above the liquid surface.
    The theoretical angular dependencies of the fluorescence yield in Fig.~\ref{fig:XSW}~(c),(d)
    were calculated for the
    following values of this parameter:
    $C_{12}=0.1$ for curve~2,
    $C_{12}=0.2$ for curve~3,
    $C_{12}=0.08$ for curve~4 and
    $C_{12}=0.03$ for curve~5.

    According to these results the distinct trend in changes of XSW data from curve~1 to curve~3 can be explained by an enhancement of the fractional amount of the Ce atoms located above the liquid surface. 
    Importantly, these XSW measurements were performed, while keeping the surface pressure constant, i.e. under continuous compression. Curves~4~and~5 were recorded at markedly different conditions, namely when the minimal area was attained and the surface pressure fell down. 
    Not surprisingly, the changes in the shape of these Ce-fluorescence curves represent completely different processes – the decrease of the fractional amount of the Ce atoms located above the liquid surface.

    It should be taken into account that good agreement between the theoretical and experimental Ce-fluorescence curves could be obtained also by varying the thickness of the Ce distribution above the liquid surface.
    In such a case the changes in the angular dependence of Ce-fluorescence from curve~2 to curve~3 can be described by the increase of the thickness of the Ce distribution above the liquid surface;
    whereas the changes of Ce-fluorescence curves 3--5 clearly indicate that this parameter gradually decreases.

    Thus, XSW data
    provided a clear idea about the Ce-induced processes in collapsed AA monolayer at $T~=~21^\circ$C. 
    During first 5 hours, when the AA monolayer was held at constant surface pressure, 
    a pronounced changes in monolayer reorganization can be attributed to
    an increase in the thickness of the Ce distribution above the liquid surface or/and an enhancement of the fractional amount of the Ce atoms located above the liquid surface. 
    During further 19 hours of measurements, 
    when the through barrier reached its uttermost position, 
    the events occurring in collapsed AA film were reversed. 
    In particular, the thickness of the Ce distribution above the liquid surface decreases or/and the fractional amount of the Ce atoms located above the liquid surface reduces.
    
\section{Discussion}
    Let us now look at all of the above observations from the perspective of the molecular organization of the monolayer.
    First, note that the AA monolayer formed on a Ce$^{3+}$ solution is thermodynamically very different from the monolayer formed on pure water, 
    as shown by the $\pi$-$A$~isotherms in Fig.~\ref{fig:iso}.

    We performed GID measurements on the monolayers with gradually increasing structural complexity. 
    First, we consider the monolayer in the solid phase.
    Compared to the control series with pure water subphase,
    the monolayer formed on a subphase with Ce$^{3+}$ ions has new structural features. 
    The Ce$^{3+}$ ions form an inorganic superstructure underneath the monolayer: 
    a periodically arranged inorganic layer with a $3 \times 2$ unit cell. 
    This is evidenced by the peaks in range A. 
    These peaks are higher than those of the organic structure, with full width half maximum in the vertical direction $\delta_z = 0.8$~\AA{}$^{-1}$, 
    from which the vertical size of the structure $d_z \sim 8$~\AA{} can be estimated to be smaller than that of the AA hydrocarbon chain $d_z \sim 22.5$~\AA{}. 
    The XSW data are insensitive to lateral order, but unlike the GID, 
    they are sensitive to the position of the Ce layer in the vertical direction. 
    The fluorescence intensity curve calculated for the Ce layer underneath the AA monolayer is in good agreement with the data in Fig.~\ref{fig:XSW}(a).
    This calculation is made for the 5~\AA{} thick Ce layer. This is in agreement with the estimation using the GID.
    Taken together these results imply that the Ce layer is indeed underneath the organic monolayer, 
    and it does form a laterally periodic inorganic superstructure.
    This is also consistent with the analysis in~\cite{miller2016observation, miller2017atomic}.
    Regarding the chemical composition of the inorganic superstructure,
    we assume that no nitrate ligands are present in the nearest environment of cerium ions in Ce(NO$_3$)$_3$ salt solution at the concentration used in our experiments. 
    This assumption is based on the results reported in~\cite{yaita1999structural},
    where X-ray absorption spectroscopy is used to study the interaction of Ln$^{3+}$ ions with nitrate ligands in aqueous solutions of Ln(NO$_3$)$_3$ salts.
    The formation of an inorganic superstructure has an effect on the structure of the monolayer itself.
    Compared to the monolayer on a pure water subphase,
    it is more densely packed with 2\% less area occupied by a molecule.
    This is apparently realized by the NN tilt of the AA hydrocarbon chains.

\begin{SCfigure}[][t!]
        \centering
        \includegraphics{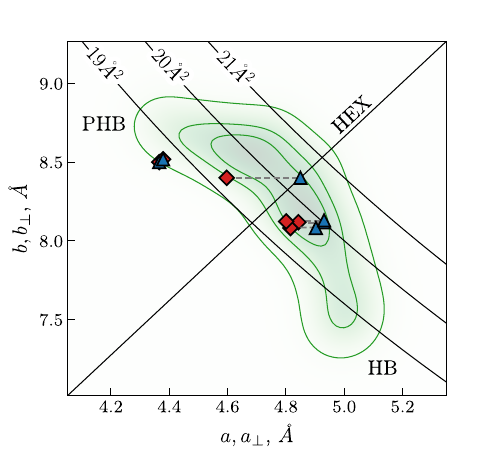}
        \caption{The $(a,b)$ lattice constants for the monolayers considered in this article. 
        The contour plot shows the arc-shaped distribution of $(a,b)$ for amphiphilic Langmuir monolayers reported in other studies.
        This distribution is assumed based on data
        from~\cite{kuzmenko1998packing,bringezu2002generic,bringezu1998influence}.
        The triangles represent the $(a,b)$ parameters from the Table~\ref{tab:structure}.
        The diamonds are the corresponding $(a_\bot,b_\bot)$ parameters of the same unit cells projected onto the plane perpendicular to the hydrocarbon chains of a fatty acid.}
        \label{fig:ab_diag}
    \end{SCfigure}

    Another structural feature is revealed in the similarly prepared monolayer formed at the lower temperature $T = 21^\circ$C: 
    the packing mode characterized by the PHB hydrocarbon chains ordering. 
    At higher temperatures, the molecular chains are in the free rotatory state. 
    The hydrocarbon chain have a flattened shape. 
    At lower temperatures, 
    the motion of the molecules is reduced and the shape of the chains begins to play a role. 
    It is then possible to reduce the area occupied by a molecule by aligning the hydrocarbon chains of neighboring molecules. 
    When the azimuth angle between adjacent chains is 90$^\circ$, 
    it is a herringbone phase mode, 
    and when it is around 40$^\circ$, 
    it is a PHB. 
    It is illustrative to mention the following. 
    Consider a plot with $a$ and $b$ lattice parameters as axes. 
    A point on this plot would represent a monolayer with certain lattice. 
    Also consider the parameters $(a_\bot, b_\bot)$, 
    which are the projections of the unit cell onto the plane perpendicular to the molecular chains. 
    In other words, 
    the parameters $(a_\bot, b_\bot)$ represent the shortest distance between molecules and are therefore more relevant when considering lattice energy calculations. 
    According to the literature~\cite{kaganer1999structure} that monolayers of amphiphilic molecules in the solid phase lie on an arc in this $(a_\bot, b_\bot)$ plot. 
    Fig.~\ref{fig:ab_diag} shows the contour plot of the distribution of different studied monolayers, 
    it is based on the data published in~\cite{kuzmenko1998packing,bringezu2002generic,bringezu1998influence}.
    Moreover it is known~\cite{kuzmenko1998packing,buzano2000herringbone} that the upper left corner of this plot is associated with PHB. 
    In Fig.~\ref{fig:ab_diag} all the structures considered in our study [see also Table~\ref{tab:structure}] are marked with triangles and their corresponding transverse $(a_\bot, b_\bot)$ parameters are marked with diamonds.
    The structures we observed in our study also obey this phenomenological rule: 
    all the diamond markers in Fig.~\ref{fig:ab_diag} are on this ark.
    Note that the solid phase monolayer formed at $T = 21^\circ$C is in the upper right corner of this plot.
    In addition, the PHB phase mode is confirmed by the presence of the (2/3 0) peak in the GID pattern. 
    It should also be noted that the structure of the monolayer has changed.
    Immediately after compression, the monolayer had a structure similar to that of the monolayer at $T = 23^\circ$C. 
    However, with time, 
    a PHB phase mode appeared. The PHB mode coexisted with the NN tilted phase mode. 
    In the PHB, the hydrocarbon chains are not tilted because all corresponding peaks are on the horizon. 
    In Fig.~\ref{fig:ab_diag} it is shown as follows: 
    the triangular marker corresponding to this structure lies exactly on the diamond, 
    as $(a,b) \equiv (a_\bot,b_\bot)$.

    %

    % {(\it A paragraph about commensurate superstructure)}

    Our experimental observations of collapse mode~1 are consistent with previous knowledge of this phase. 
    Namely, once the monolayer is compressed beyond the collapse point, 
    the 2D monolayer begins to reorganize into an ensemble of 3D aggregates on the liquid surface. 
    Indeed, the 3D aggregates are directly visible in the BAM images in Figs.~\ref{fig:bam}(b),~\ref{fig:bam}(c). 
    The implication of this process is that the lateral translational order of the monolayer decays. 
    This is observed in the GID measurements in Fig.~\ref{fig:series3},
    where we see a gradual decay of both the GID pattern of the monolayer and the GID pattern of the inorganic superstructure.

    It is quite the opposite for the type of structural ordering we refer to as collapse mode~2.
    The collapse mode~2 was observed at $T = 21^\circ$C.
    First, the BAM images of collapse mode~2 exhibit a mosaic-like structure [see Figs.~\ref{fig:bam}(e),~\ref{fig:bam}(f)].
    Second, we were able to observe the GID pattern over all 24~hours of GID measurements that implies that the monolayer retains translational order.
    According to our GID experiments, once the monolayer is beyond the collapse point, 
    it initially retains the NN packing mode, 
    further NN coexists with the PHB, and then only the PHB packing mode remains. 
    See the evolution of the GID pattern in Fig.~\ref{fig:series4}; note that the patterns in Figs.~\ref{fig:series4}(a),~\ref{fig:series4}(b) and in Figs.~\ref{fig:series4}(c),~\ref{fig:series4}(d) are measured on different samples, 
    but all controllable parameters were the same.
    The PHB packing mode is maintained for an extended period of time once collapse mode~2 is reached. 
    Although the patterns in Figs.~\ref{fig:series4}(c),~\ref{fig:series4}(d) can clearly be attributed to the PHB, since their $Q_\parallel$ coordinates measured at the horizon $Q_z = 0$~\AA$^{-1}$ match those in series~2 (cf. Fig.~\ref{fig:all}), 
    the patterns have one unusual feature: the peaks are rounded. 
    We qualitatively associate these roundings with the corrugation of the monolayer. 
    Consider a monolayer with domains that are allowed to lean from being horizontal on the liquid surface.
    Unlike the usual GID pattern, which is formed by cylindrical averaging over all possible crystallite positions and thus has a rod-like shape, 
    the GID pattern for the corrugated monolayer is formed not only by this, 
    but also by averaging over all possible orientations of the domains with respect to the surface. 
    Hence the rounding of the peaks. 
    Such peaks must lie on the circle with the origin of the reciprocal space in its center, 
    which is exactly what we observed. 
    However, these peaks are not closed to complete a ring as in 3D powder diffraction.
    This is because the greater the slope of a domain from the horizontal position, 
    the less likely that position is in the ensemble.

    \begin{figure}[t]
        \centering
        \includegraphics{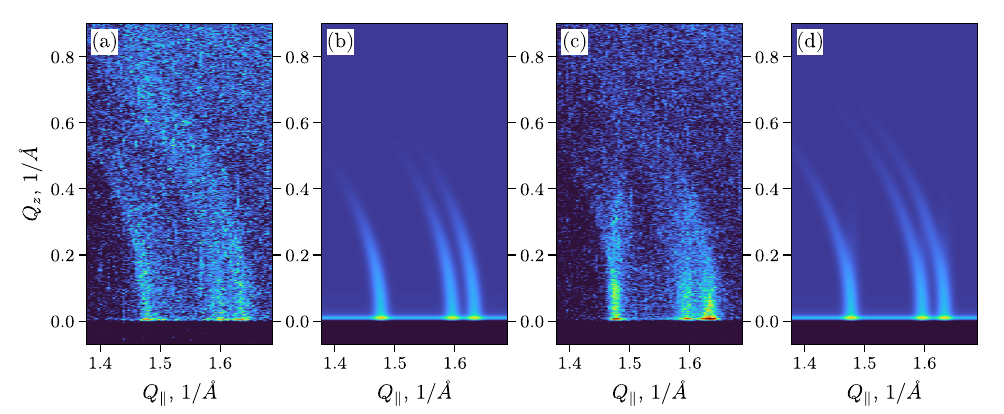}
        \caption{The GID patterns measured in series 4. 
        The start and end times of the measurement in terms of hours elapsed after monolayer compression:
        (a) 3 -- 5 hrs.
        (c) 22 -- 24 hrs.
        [data from the same sample as in Fig.~\ref{fig:series4}]
        (b,d) Corresponding best-fit GID simulations.}
        \label{fig:SIM4}
    \end{figure}

    It is of immediate interest to have a numerical characterization of such a structure.
    A simple analysis of peak positions and widths provides the orientational and translational structure,
    and also an estimate of the average size of a single crystal in the 2D powder, 
    if the resolution limit of the experimental setup allows.
    In most practical cases, such an analysis is sufficient.
    However, when analyzing partially disordered states such as collapse mode~2, 
    more information about the structure is required. This may include the parameters of the angular distribution of the domains 
    (flat single crystal regions of the corrugated film) or whether the film is partially or fully corrugated. 
    Such an analysis requires numerical simulations based on the physical principles of X-ray diffraction.
    This approach is much less commonly reported.
    In this regard, we refer to the works in~\cite{pignat2006grazing, pingnat2007grazing},
    where the effect of the structure of the fatty acid molecules is carefully considered to calculate line cuts in the 3D GID patterns.
    In~\cite{chuev2020theoretical}, an approach is proposed to calculate the full 2D GID pattern from a polycrystalline monolayer.
    Two kinds of structural imperfections are taken into account in~\cite{chuev2020theoretical}: thermal vibrations of the molecules leading to Deby-Waller type decay of the diffraction amplitude, 
    and the average size of a single crystal in the monolayer.

    We have developed a new approach for physical simulations of the GID.
    As in~\cite{chuev2020theoretical}, it is based on the distorted wave Born approximation.
    We have derived a mathematical formalism that allows flexibility in the choice of correlation functions, 
    so that different types of structural arrangements can be considered. 
    Namely, it considers the structure with long range and short range translational order. 
    More importantly for this study, it allows the perturbation of the orientational order to be taken into account.
    In addition, the new approach allows the effect of evanescent wave scattering to be taken into account, 
    which leads to the enhancement of diffuse scattering when the exit angle of the scattered ray is close to $\theta_c$.
    It also allows full 2D pattern simulations.
    This is again important for our study, since the GID pattern of collapse mode~2 has a complex configuration, 
    i.e. it is difficult to choose a set of points or a set of straight lines in the reciprocal space that fully characterizes this pattern.
    The report with the underlying mathematical model and the considered physical principles is in preparation~\cite{muftakhova2025diffraction}.

    Using the new approach, we performed the numerical simulations to reproduce the GID patterns presumably associated with corrugated structures of collapse mode~2.
    As a model of a corrugated structure, we assumed the monolayer to be an ensemble of flat crystallite domains with randomly distributed slope.
    Here, the slope is the angle between the normal to the surface and the normal to the domain.
    Furthermore, we assume that the slope is normally distributed.
    We also assume that the domains are sufficiently large so that the resulting diffuse scattering is incoherently averaged. 
    Finally, there is no preferred azimuthal orientation of the domain in our model.
    For the simulation we consider the following parameters: 
    average domain size,
    lattice parameters,
    tilt angle,
    effective thickness of the monolayer
    and 
    standard deviation of the slope $\sigma_\theta$.
    The simulations were fit to the experimental data by optimizing the model parameters using the Levenberg-Marquardt algorithm~\cite{more1978lm_algorithm}.
    
    First, we consider the GID pattern measured 5~hours after compression in Fig.~\ref{fig:series4}~(c)
    [also shown in Fig.~\ref{fig:SIM4}~(a) for reference].
    The best-fit simulation is shown in Fig.~\ref{fig:SIM4}~(b).
    Note that peak rounding is adequately reproduced in the simulation.
    In other words, the corrugated monolayer model can quantitatively reproduce the experiment.
    On average, the slope of the domain is $0^\circ$ (domain is oriented horizontally on the surface).
    This is due to the position of the intensity maximum of the rounded peaks: exactly on the horizon.
    The best fit standard deviation of the domain slope distribution [for the simulation in Fig.~\ref{fig:SIM4}(b)] is  $\sigma_\theta = 6.6^\circ$.
    This is to be interpreted as follows: approximately 70\% of the domains are inclined with a slope less then $6.6^\circ$.

    The GID pattern measured 24~hours after compression is different [see the same data in Fig.~\ref{fig:series4}~(d) and in Fig.~\ref{fig:SIM4}~(c)].
    The peaks in this pattern are partially rounded and partially straight as for the typical GID pattern of the flat monolayer.
    To simulate this GID pattern, we simply assumed that there are two coexisting phase modes: a corrugated and a flat one.
    We assumed that the diffraction on each part gives a non-coherent contribution to the final GID pattern.
    To account for this numerically,
    we introduced the ratio parameter $\gamma$ into the model.
    The parameter $\gamma$ describes the ratio of the corrugated area to the total area of the monolayer.
    
    The best-fit simulation for this model,
    with $\sigma_\theta = 8.6^\circ$.
    and $\gamma = 37\%$, is shown in Fig.~\ref{fig:SIM4}~(d).
    Again, the experimental data are reproduced:
    one can see rounded peaks together with faint straight peaks in Fig.~\ref{fig:SIM4}~(d).
    The best-fit parameters can be interpreted as follows.
    At this stage,
    63\% of the monolayer is flat,
    as in the usual case.
    The other 37\% of the monolayer is corrugated,
    but with a higher slope ($\sigma_\theta = 8.6^\circ$) than in the former case ($\sigma_\theta = 6.6^\circ$).
    To avoid confusion, we note that the $\sigma_\theta$ parameter represents the standard deviation of the slope,
    and not the slope itself.
    Such a mixed configuration of the monolayer (corrugated and flat) 
    is likely related to the fact that 5~hours after compression,
    the barrier of the trough reached its minimum position and the surface pressure of the sample was no longer maintained.
    Remarkably, the monolayer compressed beyond the collapse point maintained its structural
    ordering even 24 hours after compression.
    On the contrary, part of the monolayer became flat again.
    It is difficult to say with the data we have whether this flattening was due to relaxation of the monolayer or some other unknown self-assembly process,
    and this is an interesting topic for further study.

    These findings are confirmed by our XSW studies. 
    As discussed above the results obtained in XSW measurements implies that during first 5 hours after the compression beyond the collapse point 
    (while keeping the surface pressure constant) 
    the collapse events in AA film manifested themselves in an increase in the thickness of the Ce distribution above the liquid surface or/and an enhancement of the fractional amount of the Ce atoms located above the liquid surface.
    After that, when the area decreased to minimum value and the surface pressure fell down, 
    the thickness of the Ce distribution above the liquid surface decreases or/and the fractional amount of the Ce atoms located above the liquid surface reduces.

    A final point to be addressed regarding collapse mode~2
    is the fact that no inorganic superstructure of Ce$^{3+}$ ions was observed in the GID data.
    This observation is not consistent with the XSW measurements,
    which prove the localization of Ce$^{3+}$ ions beneath the AA monolayer.
    There are two possible explanations.
    The first is that the Ce layer is amorphous in collapse mode~2,
    so there is no diffraction.
    And the second is that the diffraction is attenuated by the distortions introduced into the Ce layer by the corrugated AA monolayer.
    In this case, one must explain why the diffraction from the AA structure is detectable while the diffraction from Ce is not.
    This may be due to the difference in effective size and shape of the scattering objects:
    AA molecules are much larger than the thickness of the Ce layer,
    and the same vertical distortions may affect the diffraction from these two structures differently.

\section{Conclusions}

    Ce-induced effects on structural reorganization during self-assembly in fatty acid monolayer were investigated depending on temperature.
    We examined the phase behavior of AA monolayer formed on the aqueous solution of Ce(NO$_3$)$_3 \, \cdot \,$6H$_2$O salt.
    GID and XSW measurements were applied to follow the changes in monolayer packing in condensed and collapsed phases at two different temperatures
    $T=21^\circ\rm{C}$ and $T=23^\circ\rm{C}$.
    The Ce$^{3+}$ ions were found to form a laterally ordered layer underneath the solid-phase monolayer in both cases.
    GID measurements revealed the formation of an inorganic superstructure of Ce$^{3+}$ ions which is commensurate with the fatty acid monolayer having a $3\times2$ superstructure.
    The localization of the Ce layer underneath the fatty acid monolayer was confirmed by XSW measurements.
    The structural organization of the fatty acid molecules in the presence of Ce$^{3+}$ ions at $T = 23^\circ\rm{C}$ was found to be very typical.
    In fact, they form a lattice in the solid phase with an NN tilt.
    The only difference is that the cell is oblique and the packing is tighter as compared to the monolayer on pure water.
    Standard collapse behavior is also observed for the fatty acid monolayer at $T=23^\circ\rm{C}$.
    In terms of GID, this is manifested by the broadening of the peaks as the structure degrades.
    The most pronounced Ce-induced effects on AA self-assembly were observed at $T=21^\circ{\rm C}$.
    In the condensed phase ($\pi=20$~mN/m) we detected the gradual transition from the NN tilted structure to the so-called PHB packing mode -- an organic $3\times1$ superstructure.
    At lower temperatures, the energy stored in the rotational motion of the molecules is reduced,
    allowing the mutual orientation of adjacent aliphatic chains,
    which in turn allows the molecular area to be further reduced,
    forming an even denser monolayer.

    Surprisingly, in the collapse region at $T=21^\circ\rm{C}$,
    the monolayer maintains its translational order despite being overcompressed.
    Quantitative analysis of 2D maps of diffraction scattering provided the fine details of molecular rearrangement in the collapsed AA film.
    Namely, the monolayer in collapse mode at $T=21^\circ\rm{C}$ consists of 2D crystalline flat domains, which are inclined from the horizontal position.
    The crystalline structure of a domain corresponds to the PHB packing mode.
    The domains are inclined according to a normal distribution.
    We deduced this by fitting forward simulations based on the DWBA to the GID data.
    Importantly, the GID pattern did not decay with time.
    The observed changes in GID can be attributed to a change in the distribution of domain inclination, while the translational order, i.e. the crystalline structure of the domains themselves, remained unchanged.
    Note that collapsed structures are exceptionally challenging objects for GID studies. 
    Nevertheless, even in such complicated cases the careful quantitative analysis of the full 2D GID pattern can provide information about fine details of mechanisms governing the collapse processes. 
    
\section*{Materials and methods}

{\it Formation of fatty acid monolayer on liquid surface.}
    Ce(NO$_3$)$_3 \, \cdot \,$6H$_2$O
    (99.99\%)
    was purchased from LANHIT Ltd and used as received.
    Arachidic acid was purchased from Sigma.
    To form a monolayer,
    a solution of arachidic acid in chloroform
    (concentration of 0.59~mg/ml)
    was spread on the surface of
    20~mM Ce(NO$_3$)$_3\, \cdot \,$6H$_2$O
    aqueous solution of pH~5.5.
    Chloroform was allowed to evaporate for 15~minutes,
    after that the layer was compressed to a desired surface pressure.
    Cerium nitrate solutions were prepared using ultrapure water
    (Millipore Milli-Q Advantage A10).

{\it Surface preasure-area measurements.}
    The Langmuir films were studied on a Minitrough Extended Langmuir trough (KSV) with a maximum interfacial area of 558~cm$^2$. 
    The compression rate was 3.75~cm$^2$/min. 
    External vibrations of the Langmuir trough were damped by an active vibration damper (Accurion). 

{\it Brewster Angle Microscopy.}
    BAM images were obtained using a Brewster angle microscope (model BAM-300, KSV NIMA) coupled to a temperature-controlled Langmuir trough. 
    A polarized He-Ne laser (wavelength 632.8~nm) was used as the light source, with the beam directed at the air/water interface at an angle of $53.1^\circ$ (the Brewster angle for water). 
    The reflected beam was collected with a 10$\times$ objective; images were taken with a USB camera and corrected with a KSV NIMA software package. 
    Measurements were performed at 21$^\circ$C and 23$^\circ$C.

{\it  Synchrotron measurements.}
    GID and XSW studies were performed at the bending magnet beamline LANGMUIR (Kurchatov Center for Synchrotron Radiation, Moscow, Russia), 
    which is equipped with a customized Langmuir trough. The closed Langmuir trough was flushed with water vapor saturated helium to reduce the scattering of the incident X-rays and to reduce the evaporation of the liquid subphase, 
    thus keeping the liquid surface position fixed. 
    The photon energy of the incident beam for all GID measurements was 13 keV. GID measurements were performed at $\theta = 0.8\theta_c$ ($\theta_c$ is the critical angle of total external reflection for water).
    The diffracted intensity in the $Q_z$ direction was recorded using a linear position sensitive detector (Mythen) with a vertical stripe size of $\mu$m. For XSW measurements,
    the fluorescence signal was recorded by an energy dispersive Vortex EX detector mounted perpendicular to the liquid surface. 
    The characteristic fluorescence spectra were recorded for each angle of incidence in the angular range corresponding to the total external reflection region.

\section*{Funding information}
    This research was funded by the Ministry of Science and Higher Education of the Russian Federation (grant no. FSFZ-2024-0003).

\section*{Acknowledgments}
    The authors are deeply indebted to Dr.~Vladimir Kaganer, Dr.~Boris Ostrovskii and Dr.~Oleg Konovalov for fruitful discussions, insightful advice, and helpfulness.

\bibliography{bibliography}{}
\bibliographystyle{ieeetr}

\newpage

\end{document}